
\input phyzzx.tex


\def\PL  #1 #2 #3 {{\sl Phys.~Lett.}~{\bf#1} (#3) #2 }
\def\NP  #1 #2 #3 {{\sl Nucl.~Phys.}~{\bf#1} (#3) #2 }
\def\PR  #1 #2 #3 {{\sl Phys.~Rev.}~{\bf#1} (#3) #2 }
\def\PRD #1 #2 #3 {{\sl Phys.~Rev.~D} {\bf#1} (#3) #2 }
\def\PRB #1 #2 #3 {{\sl Phys.~Rev.~B} {\bf#1} (#3) #2 }
\def\PP  #1 #2 #3 {{\sl Phys.~Rep.}~{\bf#1} (#3) #2 }
\def\MPL #1 #2 #3 {{\sl Mod.~Phys.~Lett.}~{\bf#1} (#3) #2 }
\def\CMP #1 #2 #3 {{\sl Comm.~Math.~Phys.}~{\bf#1} (#3) #2 }
\def\PRL #1 #2 #3 {{\sl Phys.~Rev.~Lett.}~{\bf#1} (#3) #2 }
\def\TMP  #1 #2 #3 {{\sl Theor.~Math.~Phys.}~{\bf#1} (#3) #2 }
\def\JMP  #1 #2 #3 {{\sl Jour.~Math.~Phys.}~{\bf#1} (#3) #2 }
\def\IJ  #1 #2 #3 {{\sl Int.~Jou.~Mod.~Phys.}~{\bf#1} (#3) #2 }
%
\REF\berezin{F. Berezin, {\it The Method of Second Quantization}, Academic
Press, New York, 1966}
\REF\bleher{P. ~Bleher and Y. ~Sinai, Critical Indices for Dyson's
Asymptotically Hierarchical Models, \CMP 45 247 1975 ; P.~Collet and
J. P. ~Eckmann, The $\epsilon $-Expansion for the Hierarchical
Model \CMP 55 67 1977 ;
K. Gawedzki and A. Kupiainen,
Asymptotic Freedom Beyond Perturbation Theory
(and references therein), Les Houches 1985, North Holland,
K. Osterwalder and R. Stora, Editors.}
\REF\dotsenko{V. Dotsenko and A. Polyakov, Fermion Representations
for the 2$D$
and the 3$D$ Ising Models, in {\it Advanced Studies in
Pure Mathematics}, {\bf 16}, 1988;
V. Dotsenko, 3$D$ Ising Model as a Free Fermion String , \NP B285 45 1987  ;
A. Kavalov and A. Sedrakyan, Fremion Representation of the 3$D$ Ising Model,
\NP 285 264 1987 ; P. Orland, Strings in the 3$D$
Ising Model, \PRL 59 2393 1987 ; J. Distler, A Note on the 3$D$ Ising Model
as a String Theory, preprint PUPT
1324. }
\REF\griffiths{R. Griffiths, Correlations in Ising Ferromagnets,
\JMP 8 478 1967  and 487 , \CMP 6 121 1967  .}
\REF\gross{J. Gross and T. Tucker, {\it Topological Graph Theory},
Wiley, New York, 1987.}
\REF\kadanoff{L. Kadanoff and H. Ceva, Determination of an Operator
Algebra for
the Two-Dimensional Ising Model, \PRB 3 3918 1971 ; D. Merlini
and C. Gruber, Spin 1/2 Lattice Systems , \JMP 13 1822 1972 ;
R. Balian, J.M. Drouffe and C. Itzykson, Gauge Fieds on a Lattice II,
\PRD 11 2098 1975 . }
\REF\kramers{H. Kramers and G. Wannier, Statistics of the Two-Dimensional
Ferromagnets, \PR 60 252 1941 . }
\REF\kasteleyn{P. Kasteleyn, Graph Theory and Crystal Physics, in {\it Graph
Theory and Theoretical Physics}, F. Harary, Editor, Academic Press,
New York 1969.}
\REF\geneva{Y. Meurice, Ultrametricity and Recursive Methods in
Field Theory, in
Proceedings of the International
Lepton-Photon Symposium 1991, $p$.114. }
\REF\preprint{Y. ~Meurice, Remarks Concerning Polyakov's Conjecture
for the 3$D$
Ising Model and the Hierarchical Approximation, Univ. of Iowa Preprint. }
\REF\polyakov{A. ~Polyakov, String Representations and Hidden Symmetries
for Gauge Fields, \PL 82B 247  1979 . }
\REF\samuel{S. Samuel, The Use of Anticommuting Variables in
Stat. Mechanics, \JMP \hfil \break 21 2806 1980 ;
E. Fradkin, M. Srednicki and L. Susskind, Fermion Representation of
the $Z_2$ Lattice Gauge Theory in 2+1 Dimensions, \PRD D21 2885 1980 ;
C. Itzykson, Ising Fermions, \NP 210 477 1982 .}
\REF\wegner{F. Wegner, Duality in Generalized Ising
Models, \JMP 12 2259 1971 . }
\REF\wilson{K.~Wilson, Renormalization Group and Critical
Phenomena , \PRB 4 3184 1971 .}

\Pubnum={UIOWA-91-26  }
\titlepage
\title{Duality in Long-Range Ising Ferromagnets}
\author{Yannick Meurice}
\address{Department of Physics and Astronomy, University of Iowa,
Iowa City, Iowa 52242, USA}
\vfil
\abstract
It is proved that for a system of spins $\sigma _i = \pm 1$ having
an interaction
energy $-\sum K_{ij} \sigma _i \sigma _j $ with all the $K_{ij}$ strictly
positive,
one can construct a dual formulation by associating a dual
spin $S_{ijk} = \pm 1$ to each
triplet of distinct sites $i,j$ and $k$. The dual interaction energy reads
$-\sum _{(ij)} D_{ij} \prod _{k \neq i,j} S_{ijk}$
with $tanh(K_{ij})\ = \ exp(-2D_{ij})$, and it is invariant under local
symmetries.
We discuss the gauge-fixing procedure, identities relating averages of
order and
disorder variables and
representations of various quantities as integrals over Grassmann variables.
The relevance of these results for Polyakov's approach of the 3D Ising model
is briefly discussed.
\endpage
\chapter{Introduction, Basic Ideas and Main Results}

The order-disorder duality\refmark{\kramers ,\wegner ,\kadanoff}
has been a useful tool to approach
statistical or field
theory models. In particular, the self-duality\refmark\kramers of
the two-dimensional Ising model with
nearest neighbor interactions on a square lattice
provides an easy way to
obtain the critical temperature  and
suggests a convenient way to write the partition function and the
specific heat.
The model dual to the three-dimensional nearest
neighbor Ising model is a gauge theory\refmark\wegner and
interesting loop equations, combining the order and
disorder variables, have been obtained by
Polyakov.\refmark\polyakov
These equations suggest that the model near criticality can
be described in terms
of a string model. Despite interesting results,\refmark\dotsenko
several aspects of this idea await clarification.

Recently, we have proposed\refmark\geneva
to study related problems in the case of
the hierarchical model. For this model, the
renormalization group procedure can be reduced to
a local recursion relation which has been
studied in detail\refmark\bleher and rigorous results concerning its
fixed points
and the spectrum of its tangent map have been obtained.
This recursion relation is closely related to the the ``approximate
recursion formula'' used by Wilson\refmark\wilson to estimate (rather well)
the critical exponents of the three-dimensional Ising model.
This motivated us to try to express the renormalization group transformations
in various reformulations of the hierarchical model.
In particular, we constructed the dual formulation
and the several representations of the partition
function in terms of Grassmann variables.
Due to the non-locality of the model - the price
to pay for the locality of the renormalization group transformation -
appropriate methods are necessary to enumerate the terms
appearing in the high (or low) temperature expansion. The methods we found
apply to a larger class of Ising models than the one considered
initially and are reported here.

In this paper, we consider the class of systems of spins $\sigma _i=\pm 1$,
where
$i$ runs from 1 to $N$,
with an interaction energy
$$\beta H =- \sum_{i<j} K_{ij} \sigma _i
\sigma _j \eqno(1)$$
and ${\it all}$ the $N(N-1)/2$ couplings $K_{ij}$ ${\it strictly}$ positive
but otherwise unspecified.
This class of models includes for instance, the hierarchical model
and one-dimensional Ising models
with $|i-j|^{-\alpha }$ interactions.

As usual, we define
$$ Z=\sum_{\{ \sigma _i = \pm 1 \} }
e ^{-\beta H}\eqno(2) $$
and
$$<\sigma _{i_{1}}.......\sigma _{i_{2n}}>=
Z^{-1} \sum_{\{ \sigma _i = \pm 1 \} }
\sigma _{i_{1}}.......\sigma _{i_{2n}} e ^{-\beta H}\ .\eqno(3)$$
In the following, a sequence of indices is understood as strictly ordered
unless
specified. For instance, the notation $K_{ij}$ implies $i<j$. We use
parenthesis
when the ordering needs to be performed. For instance, $K_{(ij)}$ is equal
to $K_{ij}$ if $i<j$ and to $K_{ji}$ if $j<i$. This ordering is a matter
of notational convenience and has no intrinsic significance in the problem.
We also define the dual couplings by the relation
$$tanh K_{ij}=e^{-2D_{ij}}\ . \eqno(4)$$
It is clear that the strict positivity of $K_{ij}$ implies
the strict positivity of $D_{ij}$.

One of the main results obtained here
is a dual formulation of the class of models
defined above. The disorder variables are attached to each triple of distinct
sites $ijk$ and denoted $S_{(ijk)}$. The interaction energy
can be constructed out
of products of $S_{(ijk)}$ with $i$ and $j$ fixed and $k$ running over
all its values except for $i$ and $j$. More precisely:

\noindent
{\bf Theorem 1 }. {\it The class of
Ising models defined in Eqs. (1-4) admits
the  following equivalent (dual) representation }
$$Z=(\prod_{i<j} coshK_{ij})\ 2^{N-{N-1 \choose 3 }}
\sum_{\{ S_{ijk}= \pm 1\}}
e^{+ \sum\limits_{i<j} (D_{ij} (\prod\limits_{k;k \ne i,j}
S_{(ijk)} -1))} \  \eqno(5)$$
{\it and if all the pairs} $i_1 i_2$,...., $i_{2n-1}i_n$ {\it are distinct}
$$\eqalign{& Z[D_{i_1 i_2}\rightarrow -D_{i_1 i_2},.........
,D_{i_{2n-1}i_{2n}}\rightarrow -D_{i_{2n-1}i_{2n}}]=  \cr
& Z e^{+2(D_{i_1 i_2}+......
..+D_{i_{2n-1}i_{2n}})}<\sigma _{i_{1}}\sigma _{i_{2}}
.......\sigma _{i_{2n-1}}\sigma _{i_{2n}}>\ .} \eqno(6)$$
{\it In addition, for any four distinct sites i,j,k and l , the dual
 interaction energy is invariant under the simultaneous changes
of sign of} $ S_{(ijk)},S_{(ijl)},S_{(ikl)}$ {\it and} $S_{(jkl)}$.

This theorem is proven in three steps in the next section.
The first is a standard high-temperature
expansion. The second establishes a connection between
the terms of this expansion and some elements of a vector space.
The third uses the fact that the homology of a certain boundary operator
over this vector space is trivial to rewrite Eqs. (2) and (3) under the form
given in Eqs. (5) and (6).
This proof makes the local invariance mentioned at the end of the above
theorem rather transparent. It is possible to eliminate this degeneracy
by imposing an appropriate gauge-fixing condition.
An example is given in section 3.

Theorem 1 gives a procedure to express average values of the order variables
in terms of average values of the disorder variables. This procedure
can be inverted using the standard character expansion for Ising spins.
Simple identities obtained this way are shown in section 4.
In the case of short range Ising models, identities relating appropriately
chosen products of order and disorder variables can be interpreted
as a discrete version of some kind of Dirac
(or Ramond) equation\refmark{\polyakov ,\dotsenko }.
These equations are closely related to the Schwinger-Dyson equations
associated with a representation of the partition function as an
integral\refmark\samuel over (anticommuting)
Grassmann variables.\refmark\berezin
In section 5, we rewrite
the partition function of Eq. (2) as an
integral over Grassmann variables (Theorem 2).
This formulation is very compact
and the Schwinger-Dyson equations follow straightforwardly.
Note that the integral representation
is not minimal, it involves twice as many variables as we
normally need. The elimination of these auxilliary variables is
discussed at the end of the section 6 where Theorem 2 is proven.
In the conclusions, we briefly explain how these results can be used
in a treatment of the hierarchical model inspired by the work
of Polyakov, Dotsenko and other authors.

\chapter{Proof of Theorem 1}

The first step in the construction of the
dual formulation is a standard high-temperature expansion of the partition
function and the correlation functions. We shall recall it briefly in order
to introduce useful notation.
We use the character expansion
$exp(K_{ij}\sigma _i \sigma _j)=cosh(K_{ij})+sinh(K_{ij})\sigma _i \sigma _j $
and we introduce at each link $ij$ a new variable $n_{ij}$ which
takes the values 0 or 1.
These can be used to rewrite Eqs. (2) and (3) through the identity
$$\prod\limits_{i<j} (1+A_{ij})=
\sum_{\{ n_{ij}=0,1 \} } \prod\limits_{i<j} A_{ij}^{n_{ij}} \eqno(7)$$
The non-zero contributions come from the terms where all the $\sigma _i$
appear an even number of times. The high temperature expansion can thus be
written as a sum over paths were the links $ij$ appear at most once
and the number of ``visits'' at the site $i$
$$N_i=\sum\limits_{j=1}^{i-1} n_{ji}+\sum\limits_{j=i+1}^{N} n_{ij}=
\sum\limits_{j;j \neq i} n_{(ij)} \eqno(8)$$ is odd if $i$ is one of
sites of the spins appearing in the correlation functions and even otherwise.
In the following we shall use $a=b(2)$ as
a short notation for $a=b\ modulo\ 2$.
The above arguments can be summarized by the first lemma.

\noindent
{\bf Lemma 1} {\it For the class of
Ising models defined in Eqs. (1-4)}
$$\sum_{\{ \sigma _i = \pm 1 \} }
\sigma _{i_{1}}.......\sigma _{i_{2n}} e ^{-\beta H}=
(\prod_{i<j} coshK_{ij}) \ \ 2^N \sum_{\{ {n_{ij}=0,1 \atop N_i =
m_i (2)} \} }
e^{-2\sum\limits_{i<j} D_{ij} n_{ij}} \eqno(9)$$
{\it with} $m_i=1$ {\it if}
$i \in \{i_1,.....,i_{2n}\} $ {\it and 0 otherwise.}

The second step consists in establishing a correspondence between the terms
of the above expansion and the elements of a vector space
over the field of integers modulo 2.
Our goal is to express the condition $N_i=m_i(2)$
appearing in Lemma 1 in terms
of a boundary operator having a trivial homology.
For this purpose we introduce the following definitions.

\noindent
{\bf Definitions.}
Let $V$ be a $2^N$ dimensional vector space over the field of
integers modulo 2. To any sequence of integers $1\leq i_1 < i_2 <.....
..<i_n \leq N$, we associate an element of a basis of $V$ denoted
$|i_1 i_2 ......i_n >$. The elements of this basis corresponding to sequences
with exactly $n$ indices span a subspace denoted $V_n$.
We define the {\it boundary operator} $\partial _n : V_n \rightarrow
V_{n-1}$ for $1\leq n \leq N $ such that
$$\partial _n  |i_1 i_2 ......i_n >=
|\hat i_1 i_2 ......i_n >+|i_1 \hat i_2 ......i_n >+........+
|i_1 i_2 ......\hat i_n > \eqno(10) $$
where the hat designates a skipped index. We also define $D_n : V_n
\rightarrow V_{n+1}$ for $0\leq n \leq N-1 $ such that
$$D_n | i_1 i_2 ......i_n >
=\cases{ | 1\  i_1 i_2 ......i_n  > \ &  if $ \ i_1 >1$ \cr
                                    0 \ & otherwise .\cr }
\eqno(11) $$
The choice of the first index is as inessential as the ordering itself.

We can now establish a one-to-one correspondence
between any choice of configuration $\{ n_{ij} \}$ appearing in
the high temperature expansion (9) and the vector of $V_2$
$$|\{ n_{ij} \} >\ =\sum\limits_{i<j} n_{ij} |ij> \ . \eqno(12.a) $$
Similarly, we can denote the elements of $V_1$
$$|\{ m_i \}>\ = \sum\limits_i m_i |i> \ . \eqno(12.b)$$
{}From (10), $\partial _2 |ij> = |i> +\ |j> $. This relates in a clear way
the link $ij$ and the two sites ``visited'' $i$ and $j$.
Collecting all the terms and using Eq.(7) we obtain the following.

\noindent
{\bf Lemma 2.}
{\it The condition} $N_i=m_i(2)$ {\it can be rewritten as}
$$\partial _2  |\{ n_{ij} \} >\ = |\{ m_i \}> \ . \eqno(13)$$

In order to find the general solution of Eq.(13),
it is crucial to note that the homology of $\partial _n$
is trivial. In order to establish this result, we first prove the following.

\noindent
{\bf Proposition 1. }
{\it If} $1\leq n \leq N-1$, {\it then}
$$\eqalign{&a)\ \partial _n \partial _{n+1} =0 \cr
          &b)\ D_{n-1}\partial _n + \partial _{n+1} D_n =1 \cr
          &c)\ D_{n+1} D_n =0 \ .} \eqno(14)$$
This follows in a elementary way from the definitions (Eqs. (10)-(11)).
First, $\partial _{n+1}$ deletes one index in all the possible ways and
$\partial _n \partial _{n+1}$ deletes all possible pairs of indices
{\it twice}. Since $V$ is a vector space over the integer modulo 2, this
proves {\it a)}. Similarly, we obtain {\it b)} and {\it c)}
as a mere application of the definitions by discussing separately the cases
where the first index is 1 or is not 1.

\noindent
{\bf Corollary 1.}
{\it If} $1\leq n \leq N-1$, {\it then}
$Im \ \partial _{n+1}=Ker \ \partial _n $.

Part {\it a)} of the Proposition 1 implies that
$Im \ \partial _{n+1} \subseteq Ker \ \partial _n $.
If $|v_n>\ \in Ker \ \partial _n$ then part {\it b)} implies that $|v_n>=
\partial _{n+1} D_n |v_n > $ which proves that
$Im \ \partial _{n+1} \supseteq Ker \ \partial _n$.

\noindent
{\bf Corollary 2.}
$Dim \  Im \ \partial _{n+1} ={N-1\choose n}$.

Using that $Dim \ V_n = {N \choose n }=
Dim\ Im \ \partial_n
+ Dim\ Ker \ \partial _n$ and Corollary 1, we find $Dim \ Im \ \partial_{n+1}
={N \choose n } - Dim \ Im \ \partial_{n}$. Repeating $n$ times
we obtain,
$$Dim\ Im \ \partial_{n+1} = \sum\limits_{m=0}^{n} { N \choose {n-m} } (-1)^m
\ \eqno(15)$$
which amounts to the second corollary.

We are now ready to rewrite restricted sums like those appearing in Eq (9) as
unrestricted sums.

\noindent
{\bf Lemma 3.}
{\it If f is a function} $V_2 \rightarrow $ {\bf R}
{\it and} $|\{ m_i \} >\ \in V_1$ {\it then}
$$\sum\limits_{\{ {|\{ n_{ij} \} > \in V_2 \atop \partial_2 |\{ n_{ij} \} >=
|\{ m_i \}>} \} } f(|\{ n_{ij} \} >)= 2^{-{N-1 \choose 3}}
\sum\limits_{|v_3> \in V_3} f(|\{ n_{ij}^o \} > + \partial _3 |v_3>)
\eqno(16)$$
{\it where} $|\{ n_{ij}^o \} >$ {\it is any solution of}
$$\partial _2 |\{ n_{ij}^o \} >=|\{ m_i \} >\ \eqno(17)$$
Two arbitrary solutions of Eq.(13) differ by an element of $Ker \ \partial _2$
which is equal to $Im \ \partial _3$ by Corollary 1. Consequently, we can
write any solution as the sum of as particular solution
$|\{ n_{ij}^o \} >$ plus
$\partial _3
|v_3>$ for some $|v_3> \in V_3 $. This element of $V_3$ is
determined up to an element of
$Ker \ \partial _3= Im \ \partial _4 $, a subspace
having dimension ${N-1 \choose 3}$
by Corollary 2. In other words, if we consider now $f$ to be a function over
$V_3$, it is invariant under the ``gauge transformation''
$$|v_3> \rightarrow |v_3> + \ \partial _4 |v_4> \eqno(18)$$
with $|v_4> \in V_4$.
When taking the unrestricted sum over $V_3$, we need to
divide by the ``gauge multiplicity'' which from Corollary 2
is  $2^{N-1 \choose 3}$.
This concludes the proof of Lemma 3.

We are now in position to prove Theorem 1.
Using Lemma 3, we can replace the restricted sum over $V_2$
(i.e., over the $n_{ij})$
appearing
in Lemma 1, by an unrestricted sum over $V_3$. More precisely,
for any triples of distinct ordered sites $i,j$ and $k$,
we introduce a new variable
$d_{ijk}$ taking the values 0 or 1. A given choice of $\{ d_{ijk} \} $
defines an element of $V_3$ in an obvious generalization
of Eqs. (12).  We need to replace $n_{ij}$
by ($n_{ij}^o +
\sum\limits_{k;k \neq i,j} d_{(ijk) })\ modulo \ 2$ considered
as a real number taking the values 0 or 1. This can be conveniently
done with the substitution
$$2n_{ij} \rightarrow 1+(-1)^{n_{ij}^o +
\sum\limits_{k;k \neq i,j} d_{(ijk) } }.
\eqno(19)$$
If the $m_i$ are like in Lemma 1, we can choose
for instance, $n_{i_1 i_2}^o =......=
n_{i_{2n-1}i_{2n}}^o =1$ the others being zero, as in the particular solution
appearing in Lemma 3, provided that
all the pairs $i_1 i_2$,...., $i_{2n-1}i_n$  are distinct.
The local invariance of the new formulation has been
made clear in Eq.(18).
Defining for convenience $S_{ijk}=(-1)^{d_{ijk}}$, we obtain
Theorem 1 as stated in the introduction.

It is instructive to check that the the original model can be recovered
if the duality transformation is applied to the dual model.
Starting from Eq.(5) one can proceed as above and prove that the cohomology
of a coboundary operator (algebraically dual to the boundary operator
considered above) is trivial. The original model is then recovered easily.
Since no new concepts or results are involved in this proof, it has been
left as an exercise.

\chapter{An Example of Gauge-Fixing in the Dual Formulation}

The proof of Lemma 3 has made clear the fact that the sum over $V_3$
can indeed be rewritten as a sum over $V_3 / Im\ \partial _4 $.
In this section, we show that the gauge condition
$$D_3 |v_3 > \ = \ 0 \eqno(20)$$
is sufficient to fix a unique representative for each class of
equivalence of $V_3 / Im\ \partial _4 $.
This choice is motivated by the fact that we have studied the properties
of $D_3$ in the previous section. It is not a ``covariant" choice since
the first site plays a special role. When studying particular models,
existing symmetries may suggest more appropriate choices.

In order to show that the condition (20) specifies one and only one
element of $V_3$ among those differing by a gauge transformation (18), we
introduce a projection operator $P$ defined as
$$P = 1\ + \ \partial _4 D_3 \eqno(21)$$
{}From Proposition 1, it is immediate that $P^2=P$ and more importantly that
$$D_3 P \ =\ P \partial_4 \ = \ 0 \ .\eqno(22)$$
Given any element $|v_3 > $ of $V_3$, this implies that $P |v_3 >$
satisfies the condition (20) and that if two elements differ by a
transformation (18) their images under $P$ are identical.
This concludes the proof of the statement made above.

In terms of the coefficients $d_{ijk}$ in the $|ijk > $ basis, the
condition (20) amounts to
$$d_{ijk}=0 \ {\rm if}\  1<i<j<k \eqno(23)$$
the remaining ones ($d_{1jk}$) being unspecified by the condition.
It is easy to check that a similar result may be obtained starting
with the high-temperature expansion (9) and eliminating the
$n_{1j}$ using the $N-1$ independent restrictions on the $N_i$.
(It is clear that $\sum\limits_{i=1}^{N} N_i = 0 (2)$.)

In general, a gauge condition of the form $G|v_3>=0$ specifies
uniquely the representatives and can be enforced
by a projection operator $P_G=1+\partial _4 H_3$, where $H_3$ is an
operator from $V_3$ to $V_4$ depending on $G$, provided that
we can satisfy the consistency conditions $G=G\partial _4 H_3$
and $\partial _4 = \partial _4 H_3 \partial _4 $.

\chapter{ Simple Applications of Theorem 1}

In this section, we work out simple applications of Theorem 1.
We explain how to introduce order variables in the dual formulation
and disorder variables in the original formulation.
We define the average value of a function of the dual spins $S_{ijk}$
as the r.h.s of Eq.(5) with this function inserted in the sum over the
dual spins, divided by $Z$. We shall use the same brackets
as in Eq.(3), their content
indicating unambiguously which kind of average is considered.

We know from Theorem 1 that an insertion of $\sigma _i \sigma _j $
in an average value
amounts to
a change $D_{ij} \rightarrow -D_{ij}$ in the dual formulation.
This change can be implemented
by an insertion of $exp( -2D_{ij} \prod\limits_{k;k \ne i,j}
S_{(ijk)} )$ provided it occurs only once.
The simplest application of this procedure is
$$\eqalign{<\sigma _i \sigma _j > \ &= \ < e^{ -2D_{ij} \prod\limits_{k;k
\ne i,j} S_{(ijk)} } > \cr &\geq \ 0 \ .} \eqno(24)$$
The positivity of the average follows from the positivity of the
exponential and yields the first Griffiths inequality.\refmark\griffiths
The generalization to an arbitrary number of distinct pairs is
straightforward. An interesting example is
$$\eqalign{<(\sigma _i \sigma _j)( \sigma _j \sigma _k )
(\sigma _k \sigma _i ) >\ &= \ < e^{-2((D_{ij} \prod\limits_{l;l \ne i,j}
S_{(ijl)}) \ +(D_{jk} \prod\limits_{m;m \ne j,k} S_{(jkm)})\
+(D_{ik} \prod\limits_{n;n \ne i,k} S_{(ikn)})) } > \cr
                          &= \  1 \ \ . } \eqno(25)$$
The second equality can be seen either from the fact that the square
of the sigmas is 1 or from the invariance of $Z$, as written in Eq.(5),
under the change of sign of $S_{ijk}$.

The above examples show how to express products
of an even number of order variables in the dual formulation.
The procedure can easily be inverted in order to express gauge-invariant
products of the disorder variables within the original formulation.
The practical implementation is that $exp(-2K_{ij}\sigma _i \sigma _j)$
corresponds to $\prod\limits_{k;k \ne i,j} S_{(ijk)}$
in the dual formulation. This follows from a character expansion of
the exponential, the expression of $\sigma _i \sigma _j$ in the dual
formulation discussed above, another character expansion and the
identities $sinh2D_{ij} \ sinh2K_{ij}=1$ and $cosh2D_{ij} \ sinh2K_{ij}=
cosh2K_{ij}$.
Obviously, this implies
$$\eqalign{<\prod\limits_{k;k \ne i,j} S_{(ijk)} > \ & =
\ <e^{-2K_{ij}\sigma _i
\sigma _j } > \cr  &\geq \ 0 \ .} \eqno(26)$$
The analog of Eq.(25) reads
$$\eqalign{ < \prod\limits_{j; j\ne i}(\
\prod\limits_{k;k \ne i,j} S_{(ijk)} )> \
&= \ <e^{-2 \sum\limits_{j;j \ne i}K_{ij}\sigma _i
\sigma _j } >\cr
&=1\ \ .}\eqno(27)$$
Note that each $S_{(ijk)}$ appears twice in the above products.
Using similar methods, we obtain easily
$$\eqalign{& <\prod\limits_{m;m \ne i,j} S_{(ijm)}\prod\limits_{n; n\ne k,m}
S_{(kln)}> - <\prod\limits_{m;m \ne i,j} S_{(ijm)}>
<\prod\limits_{n; n\ne k,m}S_{(kln)}>\ = \cr
                 &    \cr
& sinhK_{ij}\ sinhK_{kl}
(<\sigma _i \sigma _j \sigma _k \sigma _l>-
<\sigma _i \sigma _j ><\sigma _k \sigma _l>)\ \ .} \eqno(28)$$
The positivity of the r.h.s is the second
Griffiths inequality.\refmark\griffiths

The above procedures can be used to write products
of order and disorder variables in both formulations.
In the case of short range Ising models, it is possible to write
suggestive identities relating average values of suitably chosen
products of this type.
A better insight concerning these equations may be obtained
by reexpressing the partition function as an integral over
Grassmann variables. Note that this representation seems also well
suited for the renormalization group method\refmark\preprint

\chapter{A Representation of $Z$ as an Integral over Grassmann Variables}

In this section, we describe a representation of the partition function as
an integral over Grassmann variables (Theorem 2).
We use this integral representation
to express average values of the type considered in the previous section and
to obtain Schwinger-Dyson equations.
The proof of Theorem 2 is given in the next section.

The basic properties of a set $\{ \Psi _{a} \}$ of
Grassmann variables can be summarized as follows.
$$\eqalign{ \Psi _a ^2=& 0 \cr
\Psi _a \Psi _b =& - \Psi _b \Psi _a \cr
d\Psi_a \Psi_b =& - \Psi _b d\Psi _b \cr
\int d\Psi _a =& 0 \cr
\int d\Psi _a \Psi _a =& 1 \ . } \eqno(29)$$
A detailed presentation on this subject can be found in
Berezin's book.\refmark\berezin
This technique has been used previously for short-range
Ising models.\refmark\samuel
In the following, Grassmann variables are denoted $\psi $ or $\chi$ with
various set of indices.
\def\psij{\psi _i ^j}
\def\psj{\psi _j ^i}
\def\chij{\chi _i ^j}
\def\chj{\chi _j ^i}
\def\meas{2^N \int [d\psi d\chi]}
\def\dpdc{\prod_{i<j}d\chij d\psij d\chj d\psj }

For notational convenience, we define
$$z(x)=\cases{cosh(x)\ {\rm if }\ N \ {\rm is\ odd} \ ; \cr
                sinh(x)\ {\rm if}\ N \ {\rm is \ even}} \eqno(30)$$
$$f(x)= {d \over{dx}} ln(z(x)) \ \ \ \ \ \ \ \ \eqno(31)$$
$$[d\psi d\chi ] \ = \ \dpdc \eqno(32)$$
The following theorem will be proven inthe next section

\noindent
{\bf Theorem 2 }. {\it The partition function of the Ising models
defined in Eqs. (1-4) admits the integral representation}
$$Z=(\prod_{i<j} coshK_{ij}) \meas  \ e^{\ \sum\limits_{i<j}
((thK_{ij}) \psij \psj-\chij \chj )} \ \prod\limits_{i=1}^N z(\sum
\limits_{j;j\neq i} \psij \chij)\eqno(33)$$

As in the previous section, we define the average of a function
of the Grassmann variables by inserting this function in the integral
representation of $Z$ given in Theorem 2 and dividing by $Z$.
Again we use the same brackets, their content raising the ambiguity.
It is easy to express the average value of products of
order or disorder variables considered in the previous section in
terms of average values of functions of $\psij \psj $.
In particular, comparing the derivatives of $Z$ with respect
to $K_{ij}$ in Eq.(2) and Theorem 2, we obtain for $i<j$
$$<\sigma _i \sigma _j >\ =\ thK_{ij} + (coshK_{ij})^{-2}
<\psij \psj >\ .\eqno(34)$$
On the other hand, by changing the sign of $K_{ij}$ in these two expressions
of $Z$, it follows that
$$<e^{-2K_{ij}\sigma _i \sigma _j }> \ = \ 1 - thK_{ij}  <\psij \psj >\ .
\eqno(35) $$
These operations can be repeated in an obvious way,
in order to obtain more involved products, provided that distinct pairs
of indices are used.

It is also possible to use the well-known fact that a constant shift does not
change an integral over a Grassmann variables to obtain useful identities.
This type of identities are often called equations of motions or
Schwinger-Dyson equations.
Let $A$ be an arbitrary function of the Grassmann variables and $\Psi $
either $\psij$ or $\chij$ with $i<j$.
Using the invariance of $<A>$ under a shift in $\Psi$, we obtain the identity
$$<({\partial \over {\partial \Psi }} [ (thK_{ij})\psij \psj -\chij \chj ]
+ f(\sum\limits_{j;j\neq i} \psij \chij ) {\partial \over {\partial \Psi }}
[\sum\limits_{j;j\neq i} \psij \chij ] )\ A +
{\partial \over {\partial \Psi }}A>
=0 \ \eqno(36)$$
The function $f(x)$ has been defined in Eq.(31). In
the case $N$ even, $coth(x)$
must be multiplied the factor $sinh(x)$ appearing in the integral before
being Taylor expanded. We have used the usual convention that
the derivative with respect to $\Psi $ anticommutes with the other Grassmann
variables.
As a simple example, for $A=\psij$ and $\Psi = \psij$ the identity reads
$$<\psij \psj >\ = \ (thK_{ij})^{-1} ( 1+
<f(\sum\limits_{k;k\neq i} \psi_i^k \chi_i^k ) \chi _i ^j \psi _i ^j >)
\ . \eqno(37)$$
Similarly, if $\Psi $ takes the values  $\psj$ or $\chj$ with $i<j$,
we obtain
$$<({\partial \over {\partial \Psi }} [ (thK_{ij})\psij \psj -\chij \chj ]
+ f(\sum\limits_{i;i\neq j} \psj \chj ) {\partial \over {\partial \Psi }}
[\sum\limits_{i;i\neq j} \psj \chj ] )\ A + {\partial \over {\partial \Psi }}A>
=0 \ \eqno(38)$$

\chapter{Proof of Theorem 2 and Comments}

In this section, we prove Theorem 2, we discuss the extension of the method
to the dual case and we discuss the elimination of the ``auxiliary"
variables $\chi$.

The proof of Theorem 2 is based on the following identity
$$\eqalign{& 1+ thK_{ij} \sigma _i \sigma _j \ = \cr
& \int d\chij d\psij d\chj d\psj \ e^{\
(thK_{ij} \psij \psj-\chij \chj )} \cr
&\sum\limits_{n_{ij} = 0,1} (\sigma _i)^{n_{ij}} (\psij \chij )^{(1-n_{ij})}
\sum\limits_{m_{ij} = 0,1} (\sigma _j)^{m_{ij}} (\psj \chj )^{(1-m_{ij})}
\  .} \eqno(39)$$
After using this identity for each link, we can rearrange inside the integral
all the terms having a common subindex $i$, i.e., a common $\sigma _i$.
Summing over the $\sigma_i $, we obtain a ``local" condition namely
that
$$\sum\limits_{j=1}^{i-1} m_{ji}+\sum\limits_{j=i+1}^{N} n_{ij}\ = \ 0(2)
\eqno(40)$$
By construction, the $m_{ji}$ and $n_{ij}$ do not appear at other sites
and their sum can be performed locally.
When $N$ is odd (even), we obtain the sum over all products
of an even (odd) number of distinct $\psij \chij $.
Using the fact that the square of a Grassmann number is zero,
we can express this sum as the $cosh$ ($sinh$) of the sum over
$j$ of the $\psij \chij $. This concludes the proof of Theorem 2.

The dual version of this construction is easy to obtain.
To each link $ij$ we assign the Grassmann variables $\psi _{ij} ^k$
and $\chi _{ij} ^k $ with $k$ running over all
possible values but $i$ and $j$.
The non-local part of the integrand can be chosen as
$$e^{\ \sum\limits_{i<j} th D_{ij} \prod\limits_{k;k\neq i,j} \psi _{ij} ^k}
 \ e^{\ \gamma\sum\limits_{i<j} \prod\limits_{k;k\neq i,j} \chi _{ij} ^k}
\eqno(41)$$
where the products are ordered in the natural order and $\gamma = (-1)^{
{(N-2)(N-3)} \over 2}$. Proceding as in the above proof, we obtain the local
part of the integrand
$$\prod\limits_{i<j<k} \ sinh ( \psi _{ij} ^k \chi _{ij} ^k
+ \psi _{ik} ^j \chi _{ik} ^j + \psi _{jk} ^i   \chi _{jk} ^i )
\ \ . \eqno(42)$$

The simplicity of the proof of Theorem 2  comes from the fact
that we were able to rearrange the $\psij \chij $ into local
products. This was the reason to introduce the auxilliary variables
$\chij $ which ``screen" the anticommuting nature of $\psij $.
Had we started with the minimal identity
$$\eqalign{& 1+ thK_{ij} \sigma _i \sigma _j \ = \cr
& \int d\psij d\psj \ e^{\
(thK_{ij} \psij \psj)}\
\sum\limits_{n_{ij} = 0,1} (\sigma _i)^{n_{ij}} (\psij )^{(1-n_{ij})}
\sum\limits_{m_{ij} = 0,1} (\sigma _j)^{m_{ij}} (\psj )^{(1-m_{ij})}
} \eqno(43)$$
we would have had to specify an ordering and
to keep a detailed bookkeeping of the minus signs
resulting from the rearrangement of the $\psij $. In this process,
we obtain non-local factors of the type $(-1)^{n_{ij} n_{kl}}$ where all
the indices $i,j,k$ and $l$ are distinct (or similar expressions with
$m_{ij}$ or $m_{kl}$).
This prevents us from performing the sums locally as in the above proof.
However, we can use the decomposition
$$(-1)^{n_{ij} n_{kl}}\ =\ {1 \over 2 } (1+(-1)^{n_{ij}}+(-1)^{n_{kl}}
-(-1)^{n_{ij} + n_{kl}})\ \ . \eqno(44) $$
Each of the terms of the r.h.s
can be decomposed into local factors.
Each time this decomposition is used, the number of terms in the partition
function is multiplied by four.
We found that the number of terms in the partition
function is at least $2^{{(N-3)(N-4)}\over 6}$. This result is in
agreement with a remark of Kasteleyn\refmark\kasteleyn and its proof
in the present formulation will only be outlined.

The lower bound on the number of terms
can be obtained by establishing a correspondence
between the non-local factors and the edge-crossings of a planar
projection (depending on the ordering) of the complete graph\refmark\gross
with $N$ vertices.
We consider a surface
on which the complete graph can be embedded without edge-crossing.
{}From Euler theorem, the genus of this surface is at least
${{(N-3)(N-4)}\over 12}$, the inequality being saturated if all the
faces were triangles.
We then note that if the $n_{ij}$ and $n_{kl}$
in (43) were replaced by sums, the identity would still be valid.
At least one non-local factor of this type corresponds to the projection
of each handle of the surface considered. This concludes the outlined proof.

\chapter{Conclusions}

The duality transformation and the representations in terms of Grassmann
variables have enlighted our understanding of
the nearest neighbor Ising
models in various dimensions. We have extended these reformulations
and their immediate applications
to the case where an arbitrary number of Ising spins have a strictly
negative interaction energy associated to any pair of spin.
No other assumption has been made on the couplings.

These results apply to the hierarchical models. For these models,
the renormalization group transformation is very simple and can
be handled satisfactorily with several methods\refmark\bleher
when the order variables
are used. The study of the continuum
limit of the reformulations presented here
using the renormalization group method
is a challenging problem\refmark\preprint.
This study is expected to shed a new light
on questions related to the 3-dimensional nearest neighbors Ising model
where the hierarchical approximation is rather good, as far as the critical
behavior is concerned. In particular, we would like to understand the continuum
limit of the reformulation as ``sums over equipped surfaces" proposed by
Dotsenko and Polyakov.\refmark\dotsenko

\ack
We thank F. Goodman, W. Klink, A. Kupiainen, P. Orland
and C. Zachos for valuable
discussions. This paper is dedicated to the memory of our friends and
colleagues Dwight Nicholson,
Bob Smith, Chris Goertz and Linhua Shan.

\refout

\bye
\end